# A lyotropic ferrocolloid (ferronematic) based on a potassium laurate/1-decanole/water ternary solution

**V. V. Berejnov,*[a,b] V. Cabuil,[a] R. Perzynski[a] and Yu. L. Raikher[b]**

[a] *Laboratory of Complex Fluid, Institute of Continuous Media Mechanic,Ural Branch of RAS, 1 Korolyov Street, Perm, 614013 Russia.*
[b] *Laboratoire Liquides Ioniques et Interfaces Chargées Unité mixte 7612 CNRS Université Paris 6, 4 place Jussieu, case 63, 75252 Paris Cedex 05*
[*] *berejnov@gmail.com*

**A lyotropic ferrocolloid is synthesized by mixing of a cationic ferrofluid and potassium laurate/1-decanol/water ternary solution. Conditions of existence of a nematic phase in this mixture, other mesophases, and their phase diagrams in the vicinity of the nematic one are obtained and characterized quantitatively. Inside of the nematic zone the lyotropic ferrocolloid becomes a ferronematic, i.e., a liquid crystal with remarkably strong magnetic properties. We found that the pH and the component concentrations dramatically affect the amount of magnetic particles stably suspended in a lyotropic carrier. Magnetization and magnetic susceptibility of synthesized ferronematics are measured depending on the concentration of dispersed magnetic particles.**

## 1. Introduction

In 1970, Brochard and de Gennes [1] having assumed existence of liquid-crystalline ferrocolloids, demonstrated theoretically that these system should possess some unique orientational properties. First synthesis of a colloidal dispersion of single-domain ferrite particles on the base of a nematic liquid crystal was attempted in [2], and since then these systems are referred to as *ferronematics*. However, the synthesis [2] did not succeed to provide a stable liquid-crystalline ferrocolloid, so that ferronematic, in the proper meaning of the term, was not obtained. Despite some progress in fabrication of ferronematics achieved in the following years [3-6], neither conditions of their existence nor systematic measurements of the ferronematic material properties have been reported until now.

As it is clear, the major problem for such a complex system is to ensure colloidal stability of the dispersed ferromagnetic material in a nematogenic phase. For thermotropic nematics [5, 7] this stability is still questionable and yet far from being realized (Author's remark of 2007). Meanwhile, lyotropic nematics (micellar systems) seem quite suitable for ferronematic synthesis [3, 4, 8, 9]. Using lyotropic nematogenic ternary solution potassium laurate/1-decanol/water we have synthesized stable ferronematics [10, 11] whose life-time is at least several months.

All throughout this article we term any stable dispersion of the colloidal magnetic particles in a lyotropic medium a *lyotropic ferrodispersion* (LFD). In a particular case where the lyotropic matrix of LFD is in a nematic state, we call such system a *lyoferronematic* (N-LFD). Note that generally, LFDs might form *lyosmectic* [12-14], *lyocholesteric* [15] mesophases as well.

Liquid crystalline properties of the ternary solution of potassium laurate/1-decanol/water abbreviated as LK/1D/$H_2O$, have been studied previously [10, 11, 16-18]. As an admixture imparting to it magnetic properties we use a cationic ferrofluid [19] consisting of the $\gamma$-$Fe_2O_3$ nanoparticles in an aqueous carrier. The obtained LFDs exists both in isotropic and in nematic states. Our method [20] provides to obtain N-LFD with the refractive index anisotropy ~ $10^{-3}$ and magnetic susceptibility up to 0.07 SI units. Synthesized LFDs are stable with respect to phase-separation, and prove to be well reproducible with respect to their orientational and magnetic properties.

The essential compositional property of LFDs in the nematic phase is the orientational coupling between the suspended ferrite nanoparticles and micelles of the lyotropic phase. Due to this mechanism, it is possible to tune the optical anisotropy of a lyoferronematic using weak (<10 kA $m^{-1}$) magnetic fields [21].

## 2. Components of a lyotropic ferrodispersion

### 2.1 Lyotropic solution

Synthesis of lyotropics based on the LK/1D/$H_2O$ composition suitable for the LFD fabrication we described in detail elsewhere [10, 11]. Our method allows to prepare uniform colorless transparent lyotropic solutions long-living in the temperature interval 10–40°C. These solutions have pH ~ 10.5 and are able to reversibly produce a highly disperse foam. Depending on temperature the lyotropics exhibit uniaxial optical anisotropy.

The lyotropic LK/1D/$H_2O$ is a micellar solution with a nematic phase localizing between 10°C and 40°C. In the concentration plane the nematic phase is bound by the intervals of 5–8.8 % *w/w* and 62–68 % *w/w* for 1D and $H_2O$, respectively. Under these conditions the micellar numerical density ranges ~ $10^{19}$ см$^{-3}$. In a nematic phase, the lyotropic solution has positive uniaxial optical anisotropy [11, 16-18]. The structure of micelles for similar lyotropic solutions was studied early using small-angle-neutron scattering at room temperature [17]. The diameter and thickness of disk-like micelles were found to be ~6.4 nm and ~2.3 nm, respectively. An aggregation number,

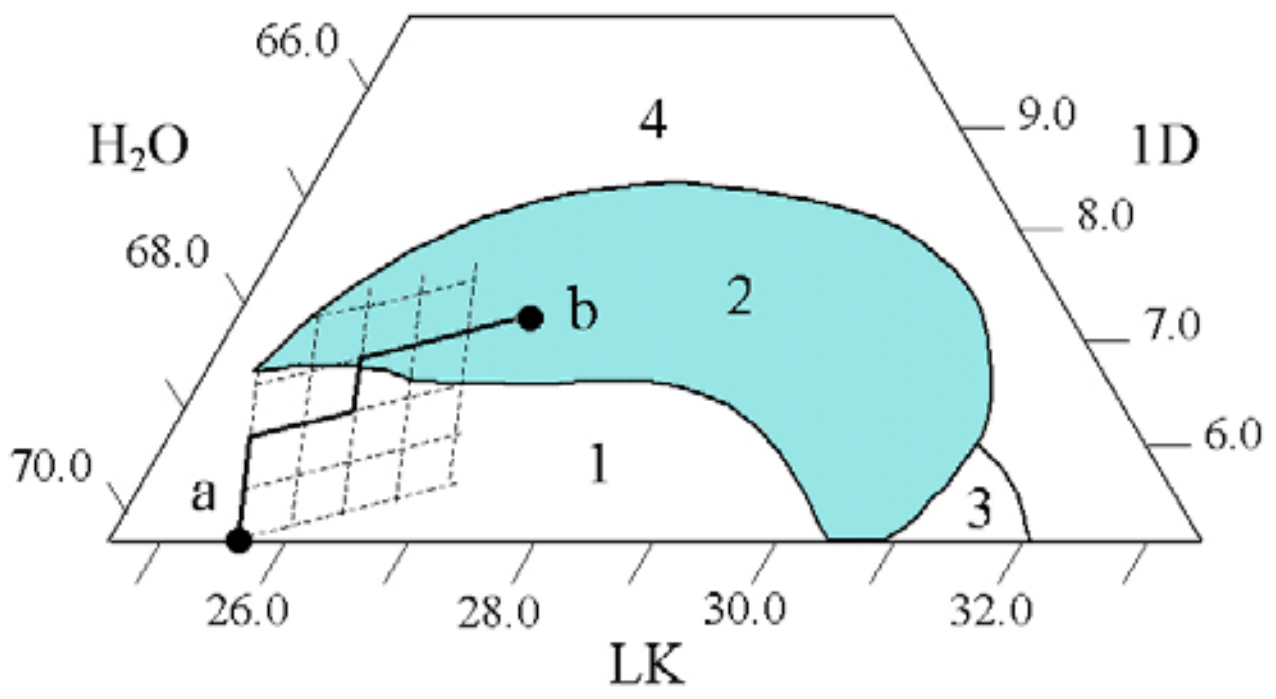

**Fig.1.** Phase diagram of a plain lyotropic solution LK/1D/$H_2O$, $T = 22^oC$, pH = 10.5. The numbers denote mesophases: (1) isotropic, (2) nematic, both of which are transparent, (3) gel-like state, and (4) crystalline state; at the boundary between phases 2 and 4 hexatic ordering occurs. Dashed straight lines represent the directions of changing the LK/1D/$H_2O$ composition by adding of $H_2O$ at $C$[LK]/$C$[1D] = const or 1D at $C$[LK]/$C$[$H_2O$] = const. The path ($ab$) is typical for lyotropic nematic fabrication.

estimated using the potassium laurate concentration, yields ~100 LK units per micelle.

### 2.2 Ferrofluid

The ionic ferrofluids were fabricated following the methods presented in [22-26]. Under standard conditions, these ferrofluids (FF) are stable aqueous colloids containing $\gamma$-$Fe_2O_3$ nanoparticles. In our samples these particles are quasi-spherical having a saturation magnetization ~ 300 kA $m^{-1}$, their size distribution is lognormal with a characteristic diameter $d_0$ = 7–8 nm, and a variance $s$ = 0.3–0.4. In Sec.4.2 we show the static magnetization curve $M(H)$ of our FF is well described by the Langevin law.

We found that cationic FF [25], which are stable in the acidic pH interval (pH=1-2) are the best to produce LFD based on the LK/1D/$H_2O$ carrier. Under synthesis conditions of the FF, the ions – $H_2O^+$ are formed at the surface of the colloidal ferroparticles (diameter ~ 10 nm) with a characteristic density ~ 200 ions per particle, while the conterions $NO_3^-$ stabilize the ferrocolloid, and are partially adsorbed at the particle surface and partially form the Debye layer in the aqueous environment close to the particle. In our FF the content of the magnetic phase varied from $10^{-3}$ up to 3 % $v/v$. No effect of the size distribution of the ferroparticles on stability of LFDs was evidenced.

## 3. Synthesis of lyotropic ferrodispersions

### 3.1 Experimental

Accurate control of weights of the components mixed in a closed container (Eppendorf or glass test tubes) at each step (Tab.1) and the following homogenization are the essential issues for the synthesis of stable LFDs. We homogenize the container content for 60–100 min (depending on viscosity) using a standard mini-vortex mixer (Vortex-Genie-2 Touch Mixer, Fisher Scientific Industries Inc., USA) and setting frequency of vibrations to 10–20 Hz. Homogenization yields a foamy content that we compact by centrifuging the container for 1 min with ~ (1–2)×$10^3$ rpm (Sigma 202 Bioblock 202, Germany).

We measured the weight of the container after the every opening (or changing the container tube) using an electronic balance and stored the data in a computer. Because of the numerous steps of

**Table 1** The procedure of LFD synthesis.

| Operation | Parameters |
|---|---|
| Control of MC* | MC volume ~9 mL |
| Addition of the LK/1D/$H_2O$ | ~3–5 mL |
| Control of MC | |
| Addition of the FF* | Drops |
| Control of MC | |
| Mixing | 10 – 20 Hz, ~100 min |
| Centrifugation | (1 – 2) $10^3$ rpm, ~1–3 min |
| Magnetic separation | ~100 kA $m^{-1}$, 10–24 h |
| Control of MC | |

MC* = mixing container, FF* = cationic ferrofluid.

mixing, the monitoring of an actual concentration of components in real time was crucial for the LFD preparation. This method allows to keep track of the composition content accordingly to the chosen phase diagram path.

During the LFD preparation, large magnetic aggregates of ferroparticles may spontaneously appear in the container. We eliminate them from the bulk of the test tube by applying a static magnetic field ~100 kA $m^{-1}$ for 10 hours.

### 3.2 Methods

#### 3.2.1 Lyotropic solution

Initially, the lyotropic solution LK/1D/$H_2O$ is prepared at room temperature in isotropic state (I) being in the close proximity to the nematic (N) transition [10, 11]. The phase diagram [10, 11, 21] of the lyotropic solution is very useful for LFD fabrication. It allows monitoring the path of "motion" of the compositional point of the ternary solution on the diagram plane during preparation. Fig.1 represents a part of the LK/1D/$H_2O$ phase diagram with a nematic domain. Note the diagram is quite sensitive to the purity of LK. The methods of preparation and purification of the latter are available in [10, 11].

When fabricating the lyotropic, it is worthwhile to "move" the composition in the phase plane using either $H_2O$ or 1D directions; these isolines are shown in Fig.1. The convenience to begin with concentrations of the lyotropic corresponding to the points in the isotropic domain 1 in Fig.1 is due the following. First, viscosity of the solutions in this domain is low that facilitates homogenization; and second, adding of 1D (up to the point where the boundaries 2-4 and 2-3 merge in Fig.1) stabilizes the previously suspended ferroparticles (see Subsec. 3.2.3). The line ($ab$) represents an example trajectory in the concentration plane for the lyotropic synthesis.

Note that before inserting ferroparticles, the lyotropic should meet the conditions stated in Sec. 2.1. Since the phase behavior of the lyotropic solution crucially depends on the pH of the composition, we maintained pH ~10.5 all throughout the LFD synthesis. The method used for pH measurement is available in [11]. It was found that the concentration of ferroparticles under which they are successfully suspended in the lyotropic composition LK/1D/$H_2O$ has a distinctive maximum at a certain pH range, see Fig.2.

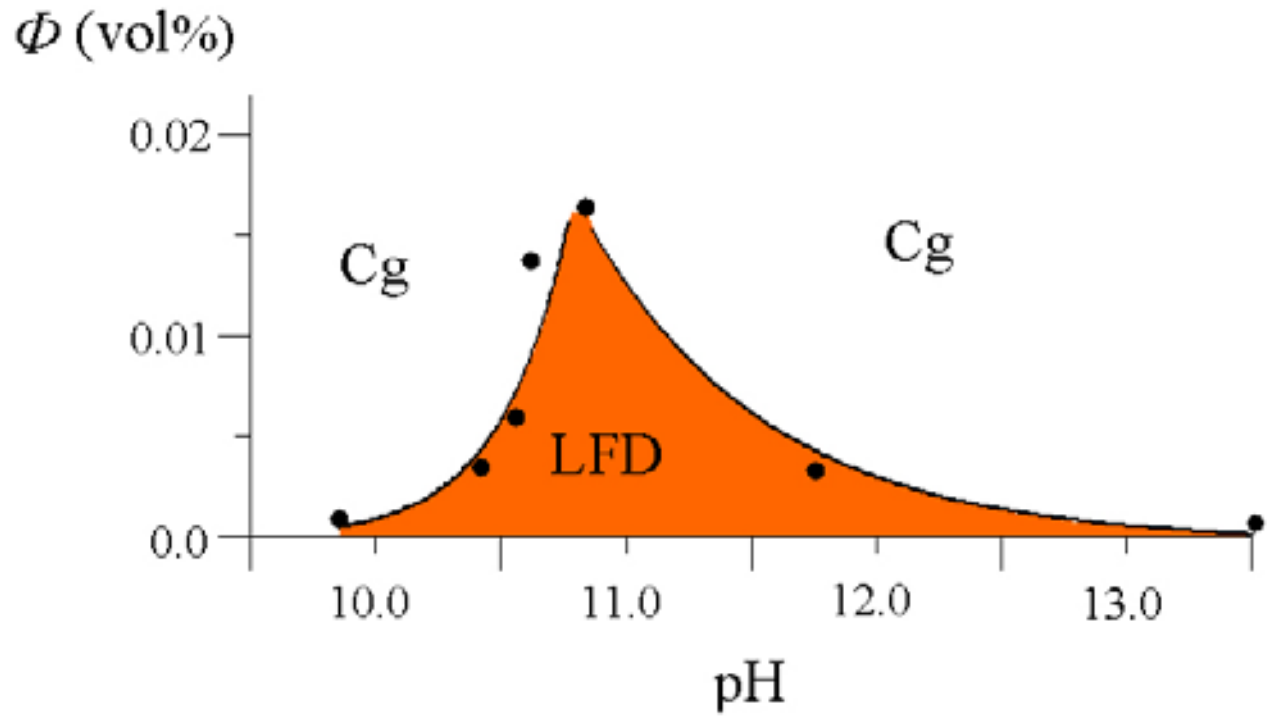

**Fig 2.** Effect of pH on the LFD stability. Composition of the lyotropic system is 27.3 % *w/w* of LK, 7.5 % *w/w* of 1D, and 65.2 % *w/w* of $H_2O$. The maximum corresponds to pH=10.7. Zones marked as Cg correspond to coagulation of ferroparticles, LFD is the zone (orange) of stable lyotropic ferrodispersion.

#### 3.2.2 Mixing of the ferrofluid with lyotropic solution

By mixing the lyotropic with an excess amount of ferrofluid we obtain a turbid resultant solution that contains LFD (the goal substance) polluted by the flakes of coagulated ferrofluid. Note that low visual turbidity should not be mistaken as a sign of instability of the LFD. We found that to adequately test for the stability of the obtained colloid LFD mixture one has to apply a magnetic field. In a stable product the ferroparticles are spontaneously suspended after the magnetic field is turned off, while in an unstable one the ferroparticles precipitate the bottom of the container. The dosage of the components for a stable LFD is determined experimentally and, as is shown below, depends on pH, temperature, and concentration.

While calculating the component concentrations of an LK/1D/$H_2O$/FF composition one should take into account that the added ferrofluid contains less then 3 % *v/v* of solid phase and, accordingly, no less than 97 % *v/v* of water. Therefore, adding a ferrofluid is largely equivalent to the change of water content, i.e., motion along the $H_2O$-axis in the phase diagram of Fig.1.

#### 3.2.3 Effect of component composition

We have found that it is namely 1D that leads to spontaneous dispersion of ferroparticles in the LFD. The sequence of images presented in Fig.3 illustrates our observation unambiguously. First, we mix a ferrofluid and a LK/$H_2O$ solution. The resultant system LK/$H_2O$/FF is heterogeneous with aggregated ferroparticles present in the form of flakes ~ 1 mm in diameter. A sample of this system was placed under a microscope, and one flake was chosen arbitrarily for further observation. Fig.3a shows the initial state of this ferroaggregate. Then a drop of 1D was added in this sample and the series of photos were registered consequently in time (Fig.3 b-d).

Fig. 4 shows the effect of 1D on concentration of suspended ferroparticles in a stable LFD for the given LK/$H_2O$ ratio. Apparently, the concentration of homogeneously dispersed ferroparticles starts decreasing at a certain concentration of 1D. Note that the maximum of concentration of the suspended ferroparticles is close to the nematic/isotropic solution transition; it decreases in the nematic (N) domain of LFD as the 1D concentration increases and reaches its minimum in the crystalline zones, Fig.1 phases (3) and (4).

#### 3.2.4 Purification of LFD

Fabricating LFD by admixing an excess amount of ferroparticles to the lyotropic carrier, results in coagulation of those ferroparticles, which a lyotropic carrier is not able to suspend even under the optimal concentration of 1D. For separating the stable and homogeneous LFD fraction from the ferro-residues we used the magnetically induced precipitation. Since the formed ferro-aggregates are large, for their elimination it suffices to use magnetic fields with the relatively low gradients ~ $10^3$ kA m$^{-2}$ applying them for ~ 24 h. Separation due to centrifugation was not successful for our LFD because at 3x$10^3$ rpm sedimentation was very slow while higher speeds destroy the LFD solutions.

### 3.3 Characterization of LFD

At the end of synthesis the LFD is a transparent and homogeneous solution with the color similar to that of usual aqueous ferrofluids. We were unable to register any ferro-aggregates by means of optical microscope. These LFD readily foam under stirring and do not separate after foam precipitation. The transition of LFD from the isotropic to nematic phase was performed by adding 1D to the isotropic LFD at room temperature. It was found that further increase of the 1D concentration after the I-N transition does not cause an increase of concentration of ferroparticles, Fig. 4.

LFD in both isotropic and nematic states could be stored for more than eight months at a constant temperature in closed testing tubes with volumes not less than 0.2 mL, Fig. 5. After magnetic purification, LFD maintains stability with respect to the gravitational field and magnetic gradients up to $10^3$ kA m$^{-2}$. However, as the volumes of LFD and the container decreases, the LFD stability also decreases. In glass capillaries (30-100 μm), both isotropic and nematic LFD separate into layers in 1–2 days. According to our observation, LFD's lifetime decreases as a ratio of the capillary wall area to the capillary volume increases.

The highest ferroparticle concentration in N-LFD reached in our experiments was ~ 1 % *v/v* (~ $10^{15}$–$10^{16}$ ferroparticles per cm$^{-3}$). However, ferronematics with a highest magnetic content are less stable. We found that ferronematics with lower ferroparticle concentration, 0.05–0.1 % *v/v* are more stable and much more fit for magnetic and optical experiments.

## 4. Magnetic properties of lyotropic ferrodispersions

### 4.1 Method

We measured magnetization of ferronematic and ferrofluid samples using a vibrating sample magnetometer. The LFD sample filled completely a glass cylindrical tube with the inner radius 3 mm, volume 1 mL, and the aspect ratio 1/5. This sample tube was placed in the pick-up coil (diameter ~ 1 cm) where it was vibrated sinusoidally at 120 Hz in the direction perpendicular to the external, uniform magnetic field. This vibration induces a sinusoidal current in the pick-up coil proportionally to the sample magnetization. Varying the uniform field from 0.5 to $10^3$ kA m$^{-1}$ within the time intervals from 10 min to 1 h we were able to measure the sample magnetization as a function of the magnetic field; all the observations were performed at 20C.

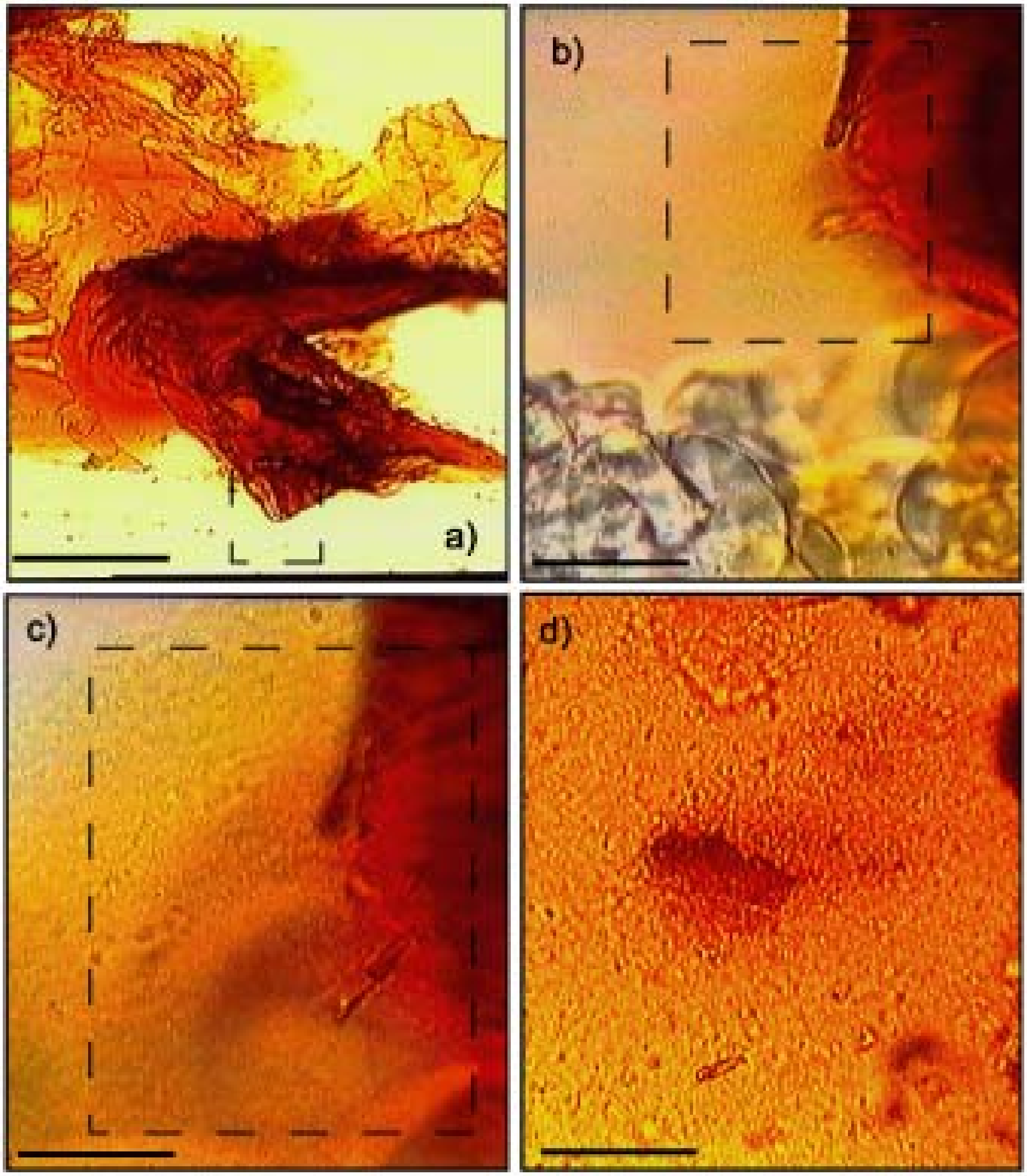

**Fig. 3.** Decanol induced re-dispersion of magnetic coagulates in LK/$H_2O$/FF. The dashed frame in Fig.3a outlines a part of one of the flakes chosen for further observations (b-d); the bar length is 1 mm. Before adding 1D, this pattern is stationary. Fig.3b represents the ferro-flake 60 min after a drop of 1D (occupies the lower part of the image) was added; the bar here is 200 μm; as seen, typical myelin fingers appear at the boundary between LK/$H_2O$ and 1D. Figs.3c and 3d correspond to 100 and 120 min after adding the 1D drop, respectively; the reference bars are 100 μm and 50 μm long.

### 4.2 Results and discussion

Fig. 6 shows magnetization of three LFD samples in the nematic phase. A simple formula $M(\infty) = \Phi I$ describes the saturation magnetization, where $\Phi$ and $I$ are the volume concentration and magnetization of the ferrophase, respectively. Noteworthy that the magnetic properties of the LFD are isotropic within the experimental accuracy and no magnetic hysteresis is visible, Fig.6. This result does not contradict the previously observed dependence of the magneto-orientational properties for thin ferronematic layers on the direction of the applied magnetic field [21]. The fact is, that the proper anisotropy of magnetic properties of N-LFD is a rather week effect, which rapidly decreases with increasing size of the sample. Indeed, according to [21], the orientation-elastic anisotropy in a 100-μm layer of N-LFD has the same effect on the magnetic particles as application of a magnetic field ~ 3 kA $m^{-1}$, which is inversely proportional to the size of the container. Therefore, the effective field induced by the liquid crystal magnetic anisotropy in the pick-up coil must not exceed 30 A $m^{-1}$ that is an order of magnitude smaller than the resolution value of our magnetometer.

For the tested LFD and their parent ferrofluids we did not find any difference in the magnetization $M$ and the initial susceptibility $\chi_0$. Fig.7 shows the normalized magnetization $M/M_0$ ($M_0$ is the saturation magnetization of LFD) as a function of the magnetic field for different concentrations of the magnetic particles. The collapse of the curves to one (master) line indicates linear dependence of the LFD magnetization on concentration of the suspended magnetic particles.

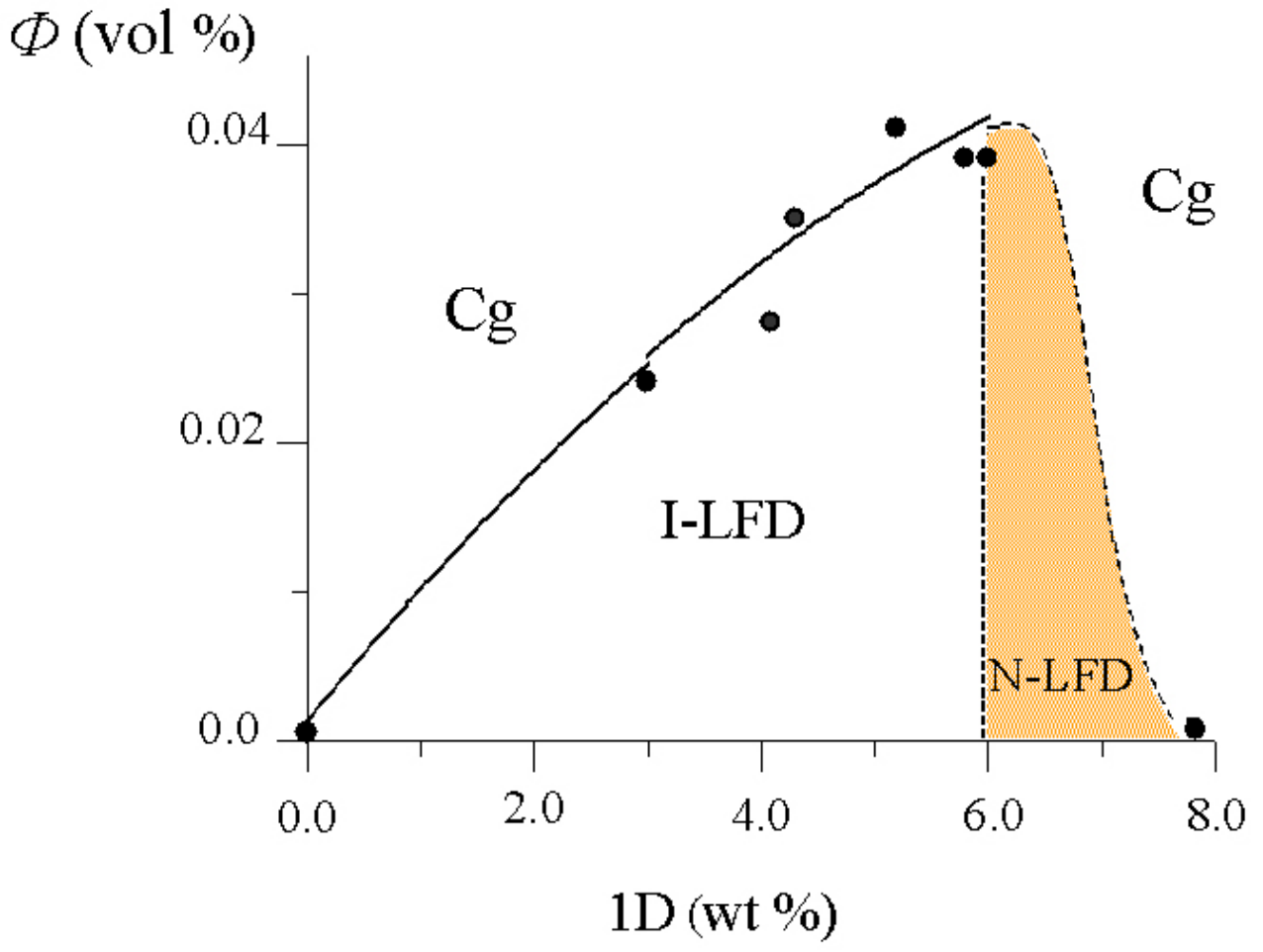

**Fig.4.** Effect of 1D on stability of LFD in isotropic (I) and nematic (N) phases. The concentration ratio $C[H_2O]/C[LK]$ = 2.31. Zone Cg corresponds to coagulation of ferroparticles; in the zones I-LFD and N-LFD the ferroparticles are well dispersed in the lyotropic carrier, these LFD are stable and transparent.

A solid line presented in Fig.7 is a result of theoretical calculation within the assumption that the ferroparticles are polydisperse and their magnetization is superparamagnetic (the Langevin law). The distribution of ferroparticles in this calculation was assumed to be lognormal with the parameters $d_0$ = 8.2 nm and $s$ = 0.3, which were obtained due to fitting the experimental magnetization curves of the parent ferrofluids under different concentrations. A fairly good agreement between the LFD data and the theoretical calculation in Fig.7 means that the distributions of ferroparticles in those colloids are equal. Thus we conclude that the developed method of LFD fabrication does not change the distribution of ferroparticles in comparison with the parent ferrofluid.

One might expect that the magnetic susceptibility $\chi_0$ linearly depends on volume fraction $\Phi$ of the magnetic particles for the dilute (< 1 % *v/v*) LFD. The data presented in Fig.8 confirms this assumption yielding $\chi_0 \approx 7.3\ \Phi$ that amounts to $\chi_0 \approx 0.07$ SI in the case of maximal ferroparticle concentration (~ 1 % *v/v*) achieved in our LFD.

The average particle size may be estimated using the slope $(\chi_0/\Phi)$ in Fig.8. Indeed, for non-interacting Langevin superparamagnetic particles the initial susceptibility reads $\chi_0 = \mu_0 I^2 V\Phi/3k_BT$, where $\mu_0$, $I$, $V$, and $k_BT$ are the magnetic constant, magnetization of the ferroparticle material, particle average volume, and temperature in energy units, respectively. The average particle diameter $d$ we define then assuming $V = (\pi/6)d^3$. Then the preceding formula yields $d = [(18k_BT/\pi\mu_0 I^2)\ (\chi_0\ /\Phi)]^{1/3}$. Substituting $I$ = 310 kA $m^{-1}$ and the ratio $\chi_0/\Phi$ = 7.3 from the plot of Fig.8, we obtain $d$ = 11.5 nm at room temperature. This estimate agrees well with the $d = d_0\exp(4.5s^2)$ = 12.2 nm that one obtains using the lognormal law with the above-established parameters of the ferroparticle distribution.

Note that magnetic fields that cause the macroscopic optical effects occurring in the 30–500 μm ferronematic layers are less than ~10 kA $m^{-1}$ [21]. This involves just the initial part of the magnetization curve, where the linear law: $M = \chi_0 H$ holds Fig.6.

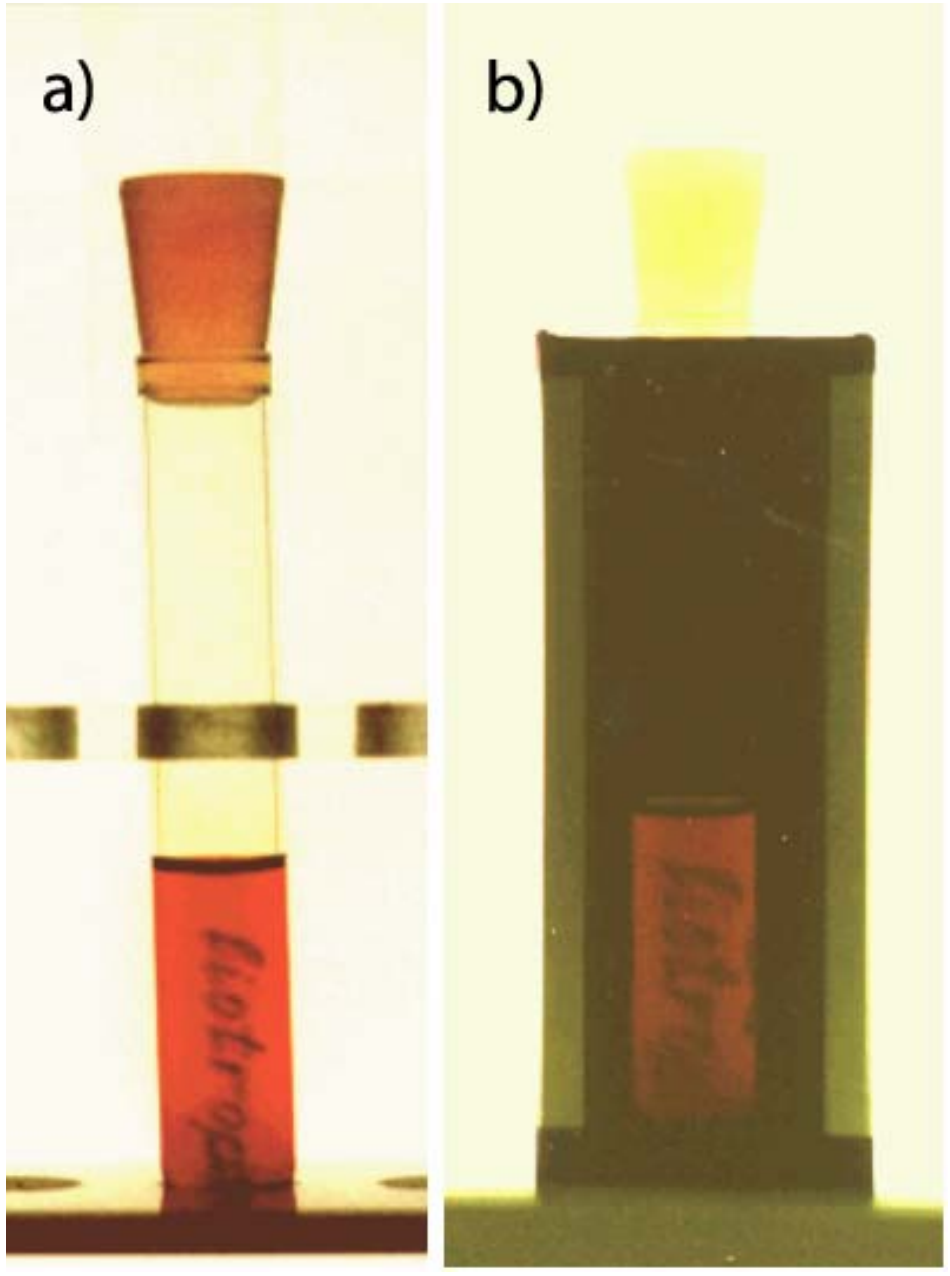

**Fig.5.** Views of a stable and transparent LFD sample in white (a) and polirized (b) light. LFD is in the nematic state; the word "liotropic" is written on a transparent screen behind the test tube..

If we place a LFD sample in a nonuniform magnetic field then ferroparticles will drift toward the higher field. This effect may disturb the local balance of magnetic particles and cause LFD phase separation. To estimate the magnitude of this effect the following experiments were performed. Cylindrical tubes of 5.5 cm length (Bioblock-PS, Germany) were filled with LFD (~ 0.06 % *v/v* of ferroparticles) and placed in a vertical gap between the poles of a permanent magnet to obtain the magnetic field perpendicularly to the tube axis. The field at the tube bottom was about $10^3$ kA $m^{-1}$ (center of the poles) whereas the free-interface of the LFD was in a rather weak field at a distance of ~ 2 cm beyond the edge of the magnet poles.

Magnetophoretic drift of the ferroparticles was observed producing distinct visual nonuniformity of the test tube color. In about 3 h, the concentration of ferroparticles near the free interface and at the bottom of the test tube became ~ 0.001 and > 0.1 % *v/v*, respectively. Although the LFD sample becomes opaque in the region of the highest ferroparticle concentration it remains stable: one does not observe any phase separation. If the test tube is removed from the magnetic field and let it be at rest under zero field for some time, the uniform particle distribution is spontaneously restored.

## 5. Phase characterization and study of phase transitions in LFD

### 5.1 Experimental

For the phase identification and characterization we used mini test tubes (Prolabo, UK) of 1 mL with tight polyethylene lids and capillaries (VitroDynamic, UK) 0.1 mm thick x 2.0 mm wide. The capillaries were sealed thoroughly and irreversibly attached to microscopic glass slides (Esco, Portsmuth, USA). The samples were placed into a thermostat thermal equilibrium we reached in ~ 40 min and maintained with the accuracy ~ 1°C.

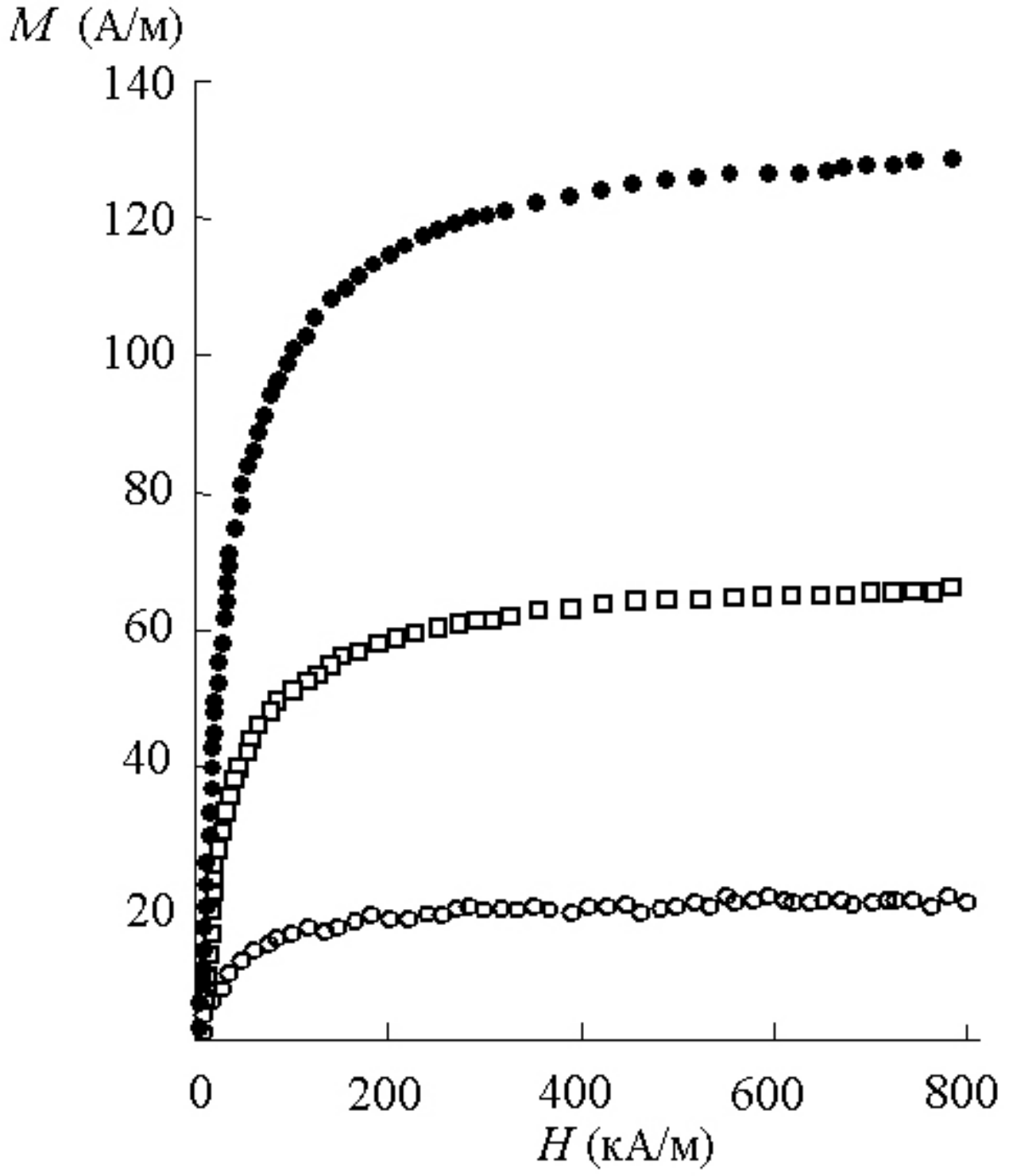

**Fig.6.** Magnetization curves of LFD. The ranges of the lyotropic components are 27–28.5 % *w/w* for LK and 5.7–7.5 % *w/w* for 1D; ferroparticle concentrations: (○) 0.008, (□) 0.028, and (●) 0.059 % *v/v*..

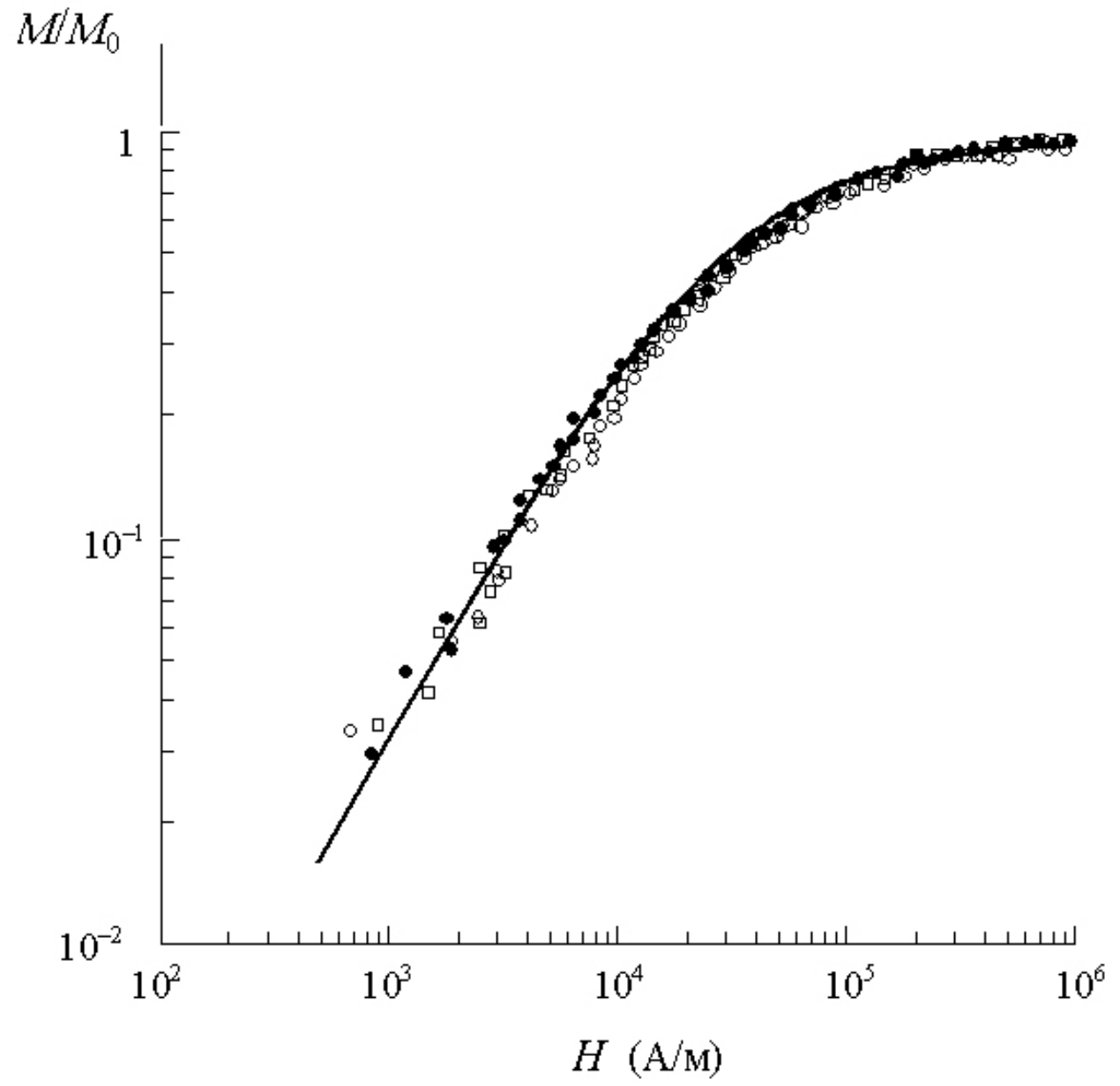

**Fig.7.** Data of Fig.6 after normalization. The solid line shows Langevin approximation taking into account lognormal distribution of ferroparticles with the parameters $d_0$ = 8.2 nm and variance $s$ = 0.3. These parameters were obtained due to fitting the magnetization curves of the parent cationic ferrofluids with Φ = 0.12 and 0.43 % *v/v* vol used for LFD fabrication; magnetization of the particle substance is set 310 kA $m^{-1}$.

We identified the liquid-crystalline and dispersion phases visually using the polarizing crystallographic microscope POLAM R-111 (LOMO, Russia). For express analysis we observed the samples through a polarization cell (a box with the two crossed polarizers inserted in the opposing walls).

The ferroparticle concentration was determined using two different techniques: measurements of LFD magnetization with the vibrating sample magnetometer and colorimetric method, i.e., visual comparison of the sample color with the color of ferrofluids, which were in advance calibrated with respect to concentration of the dispersed ferrophase (see Appendix).

A probed portion of LFD with a given LK/1D/$H_2O$/FF ratio was poured out in three identical test tubes and 4–5 capillaries. Having in possession a number of identical samples allowed us to map the phase boundary precisely even if irreversible structural transformations occurred.

### 5.2 Method

For mapping phase diagrams of LFD the method described in [10] was modified. We have chosen transparency of the solution as an additional parameter characterizing colloidal stability. If a particular LFD sample is not transparent we do not consider its state as stable even if the time of its visual separation is long. LFD phase diagrams were plotted by modifying gradually a certain selected parameter: pH, temperature, or a component concentration. At some value of the parameter, the initially transparent LFD possessing a characteristic color of a magnetic fluid suddenly becomes turbid in the all-accessible volume. This turbidity means that the sample contains more than one phase and/or large magnetic aggregates. Applying a gradient magnetic field to a turbid LFD sample, it is possible to separate the LFD phase by collecting coagulated magnetic aggregates in sediment.

Phase observations at each given temperature were performed as follows. A series of samples with different concentrations of a certain component was thermostated until thermal equilibrium was reached. Next, each sample was taken out from the thermostat for ~ 1 min and the phase was examined under a polarizing microscope. In some cases, for express analysis, a box with cross-polarizers was used. We found that disturbances of the thermal regime during observation do not cause any changes in the current phase state.

### 5.3 Phase diagram of LK/1D/$H_2O$/FF

Fig. 9 shows the phase diagram of LFD. It turns out be significantly different from the phase diagram of the initial lyotropic solution studied in [10, 11]. First, LFD exhibits lower number of phases than the pure lyotropic solution. For example, hexatic (Fig.1, (4)) and gel (Fig.1 (3)) phases [10, 11] do not exist in LFD. In particular, the ferroparticles coagulate completely, and LFD ceases to exist as a system, when one tries to reach any of the afore-mentioned phases changing either temperature or concentration.

Second, the size of the nematic domain for LFD decreases as the ferroparticle concentration increases. This effect becomes the more pronounced the closer is concentration to 1 % *v/v*, which in our case seems to be the limit.

Third, the temperature boundaries of stable phases change their places significantly in comparison with a pure lyotropic solution, [10, 11]. For LFD the interval between the upper and lower bounds is quite narrow. However, oppositely to pure lyotropics, the region of existence of stable LFD expands towards the low temperature and reaches $0^oC$ which is quite unexpected, since usual lyotropics turn into either polycrystalline or gel phases at about $13^oC$, [10, 11]. Outside of these limits, LFD are not stable because of the partial (sometimes total) coagulation of ferroparticles. This coagulation is *irreversible*: when the initial temperature of the coagulated LFD is restored, no spontaneous re-dispersing takes place. In this case, even forced re-dispersing due to mechanical or ultrasonic homogenization does not produce stable and transparent LFD.

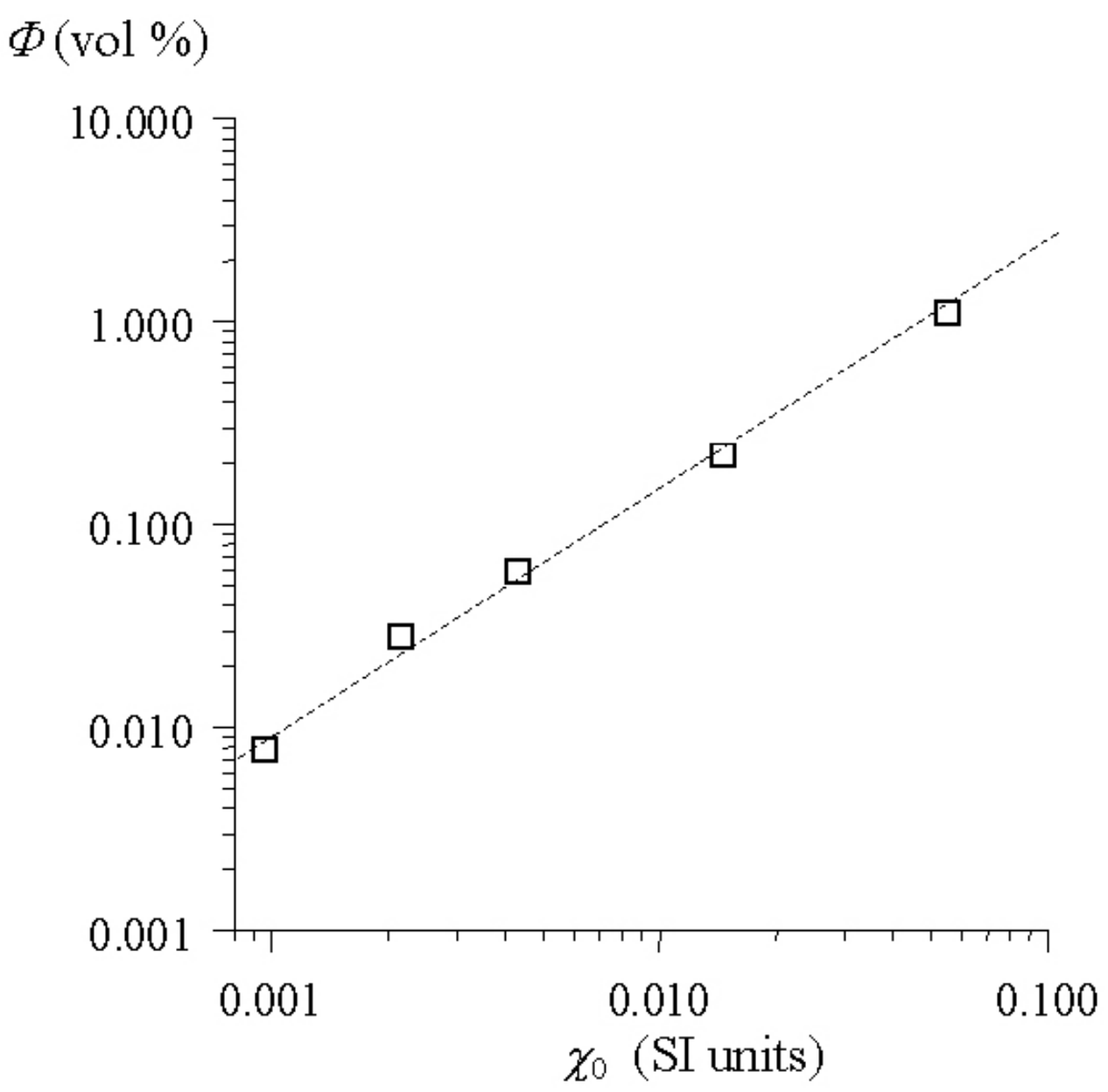

**Fig.8.** Correlation between the volume concentration of ferroparticles Φ found chemically (see Appendix) and isotropic magnetic susceptibility $\chi_0$ found from magnetic measurements. The changes in Φ cause the changes in the compositional concentrations of the lyotropic LK/1D/$H_2O$; the slope of the straight line is $\chi_0 / \Phi = 7.3$.

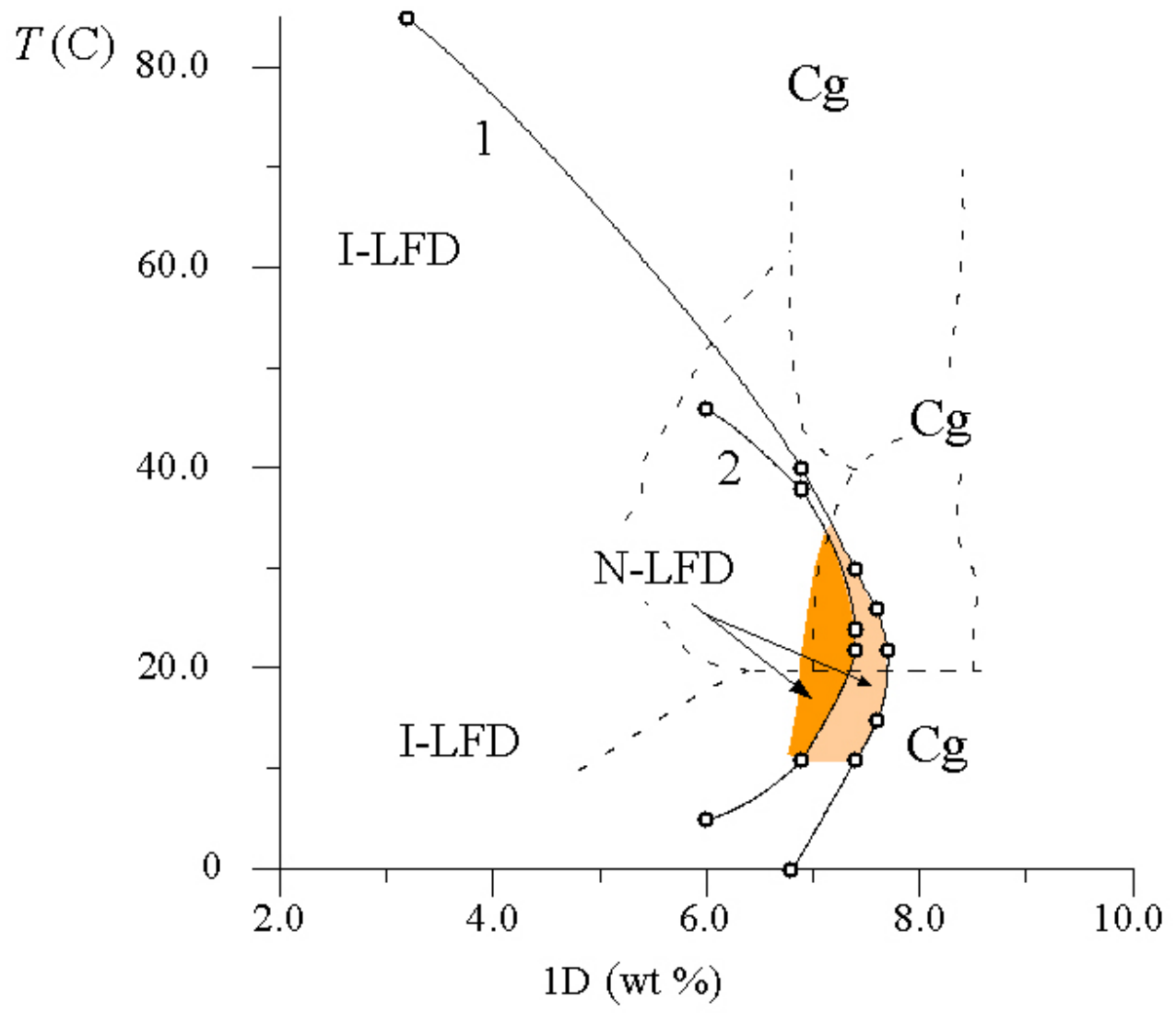

**Fig.9.** Phase diagrams of the LFDs with constant ratio $C[H_2O]/C[LK]$ = 2.51 and different ferroparticle concentrations: (1) 0.04 and (2) 0.07 % *v/v*. Dashed lines present the phase diagram of a plain lyotropic LK/1D/$H_2O$ [10, 11]. Solid curves bound zones I-LFD, N-LFD, and Cg which correspond, respectively, to a stable and transparent LFD, and to ferroparticle coagulation. The orange regions inside the zone of stability mark the nematic state; their intensity corresponds to ferroparticle concentration.

Despite the fact that it is possible to prepare N-LFD with the high concentration of ferroparticles, long-living, highly concentrated (Φ > 0.5 % *v/v*) N-LFD require special caution.

### 5.4 Phase Transitions

We observed two distinct phase transitions destroying LFD. The first one is a separation of the initially homogeneous LFD in a lyotropic solution and a magnetic sediment. This separation corresponds to appearance of uniform turbidity of the sample. In Fig.9 the solid lines marks the boundary of such a transition. The above separation is irreversible with respect to the temperature: a decrease of temperature (cooling) does not induce spontaneous re-dispersing of the aggregates. However, it is possible to successfully re-disperse those aggregates by changing concentration of 1D. The last method works only while the LFD is still far enough from the region of the hexatic phase (Figs.1, part of the phase (4) close to the 2-4 boundary).

The second transition is observed in the stability zone and corresponds to the isotropic – nematic (I-LFD – N-LFD) transformation. This transition causes the appearance of a *ferronematic*. The typical ferronematic domains are marked in Fig.9 as orange zones of different intensity. The size of the N-LFD zones decreases as the ferroparticle content increases.

It is worthwhile to note, that intense visually detectable fluctuations of the optical anisotropy appear in LFD macro-samples as the system approaches to the I –N transition. Similar effect was registered in pure lyotropics [10, 11]. It is induced by a very weak mechanical perturbation in result of which initially isotropic LFD sample turns into N-LFD and transforms back to I-LFD in a period less than a second. Putting a sample in a box with windows made of crossed polarizers, one sees a white flash in response to a gentle shaking of the cell. The size and duration time of the flash but weakly depends on the concentration of ferroparticles (up to ~ 1 % *v/v*). Applying a magnetic field ~ 10 kA $m^{-1}$ to an I-LFD sample in the state close to the I – N transition, does not provide any affect of the induced optical anisotropy while the same field is sufficient for macroscopic orientation of the N-LFD samples [21].

## 6. Conclusions

Method of fabrication of stable lyotropic ferrodispersions, LFD is developed. This method is based on proper engineering of the lyotropic solution LK/1D/$H_2O$ and a procedure of mixing of this solution with a cationic ferrofluid, that is an aqueous colloid (pH~1.5) of magnetic particles of $\gamma$-$Fe_2O_3$ with the mean size ~ 7–8 nm.
The stability diagrams of LFD are mapped with respect to the concentration of 1D, ferroparticle content, pH value, and temperature. A non-monotonic effect of 1D concentration and pH on the content of ferroparticles is established.

Thick layers (~1 cm) of the fabricated LFD are highly transparent similarly to pure lyotropic LK/1D/$H_2O$. Depending on 1D concentration and temperature the conditions of I – N transition are measured. The refractive index of the nematic phase of LFD was found to be positive that identifies the obtained N-LFD as a *discotic ferronematic*.

Stable dispersion of ferroparticles is achieved both in the isotropic and in the liquid-crystalline states. The transition I-LFD – N-LFD does not provoke phase separation of LFD. Near the I –N transition, macroscopic fluctuations causing spontaneous appearance of optical anisotropy (similar to those in a pure lyotropic solution [10, 11] were observed.

Initial magnetic susceptibility and magnetization of thick LFD layers were measured in both nematic and isotropic states. Those characteristics are found to be very similar to those of ferrofluids used for LFD fabrication. Static magnetization of LFD resembles the Langevin law of superparamagnetic magnetization. We did not find the difference in distribution of magnetic particles in LFD and the parent ferrofluid. In the concentration range studied, the initial magnetic susceptibility $\chi_0$ is linearly proportional to the concentration $\Phi$ of the suspended ferroparticles.

We found both I-LFD and N-LFD to be stable in magnetic fields up to $10^3$ kA m$^{-1}$. A non-uniform magnetic field cases a concentration gradient of ferroparticles in LFD. However, a LFD sample retains its stability and the ferroparticle stratification disappears reversibly when the magnetic field is turned off.

Yet nowadays the theoretical aspect of ferronematics is not appreciated thoroughly. Moreover, there is an evident “gap” between theory and experiment. Indeed, starting from the pioneering work [1], theory was developed [27-30] only with the assumptions: (*i*) that the liquid crystalline base of a ferronematic is a thermotropic mesogen, and (*ii*) that the ferroparticles are highly anizometric and thus posses quite a high magnetic anisotropy. In fact, up to date the only existing ferronematics are the ones constituted by a lyotropic base with superparamagnetic (low magnetic anisotropy) ferroparticles embedded therein. These differences are essential for theoretical modeling [31]. First attempts to develop a more adequate theory of lyotropic ferronematics were made in [21, 32].

Authors thank S.N. Lysenko for helpful discussions and criticisms. This work was partially supported by “Le Réseau Formation – Recherche Europe Centrale et Orientale” (grant 96P0079) and Russian Foundation for Basic Research (grant 98-02-16453).

## Appendix

### Methods of Measurements of Ferroparticle Concentration in LFD

As noted above, during the LFD synthesis we remove the coagulated magnetic particles from the stable dispersion using magnetic separation (Subsec. 3.1 and 3.4). As the concentration of the suspended ferroparticles changes, this loss should be measured.

### Magnetic method

Our magnetic measurements (Sec. 4) show that magnetic properties of LFD (under ferroparticle concentration < 1 % *v/v* and at moderate fields < 10 kA m$^{-1}$) are similar to those of their parent ferrofluids. Since $\chi_0$ depends on $\Phi$ linearly, for estimating the ferroparticle concentration in LFD it is possible to employ the method used for pure ferrofluids. Measuring magnetization of a sample of a given shape it is possible to obtain the concentration of magnetic particles by comparing this magnetization with the magnetization of another sample whose concentration is known. This straightforward method, however, requires some high-precision equipment for concentrations below 0.01 % *v/v*.

### Colorimetric method

It is based on measuring the color of LFD. Since a stable LFD sample is highly transparent and retains color of a parented ferrofluid, one can relate $\Phi$ to the color of the LFD sample. Thus we can find $\Phi$ comparing the colors of LFD and pure FF samples with known ferroparticle concentrations.

To do that, a sequence of FF samples of different colors with concentration increment $10^{-3}$ % *v/v* was prepared. Under intense illumination a trained eye can distinguish these colors difference, and hence determine ferroparticle concentrations up to 0.7 % *v/v* with resolution about $10^{-3}$ % *v/v*. The ferroparticle concentrations in FF were evaluated with the aid of the magnetic method described in Sec. 4 and by measuring molar concentration of iron in ferrofluids by a direct chemical method. When a series of colorimetric samples is prepared, the corresponding measurements take ~ 1 min.

We have checked, how much the colorimetric $\Phi$ deviates from the value of $\Phi$ calculated assuming that all the ferrofluid is dispersed in LFD in result of mixing. In the case where mixing leads to complete dispersion of ferroparticles the errors of the colorimetric and mass methods are always of the same order of magnitude.